\documentclass[aps,prl,10pt,twocolumn,superscriptaddress,preprintnumbers,floats]{revtex4}
\usepackage[dvipsnames]{xcolor}
\usepackage{physics}

\usepackage{braket}
\usepackage{enumerate}

\usepackage{scalerel}
\usepackage{blindtext}

\usepackage{latexsym}
\usepackage{natbib, comment}

\usepackage{mathrsfs,amsmath,amssymb,amsthm,amsfonts,tikz,graphicx,accents,hyperref,color}
\usepackage{url}
\usepackage{dcolumn}
\usepackage{multirow}
\usepackage{color}
\usepackage{cancel}
\usepackage{soul}
\usepackage[normalem]{ulem}
\usepackage{txfonts}
\usepackage{epsfig}
\usepackage{psfrag}
\hypersetup{colorlinks=true}
\usepackage{mathtools}
\usepackage{float}
\usepackage{caption}
\usepackage{subcaption}
\usepackage{ragged2e}

\usepackage{slashed,ccaption}

\usepackage[all]{xy}

\usepackage{multirow}

\usepackage{array}
\hypersetup{ linktoc=all,
    colorlinks, linkcolor={palatinateblue},
    citecolor={red}, urlcolor={amaranth} 
}

\graphicspath{{Images/}}

\definecolor{rosy}{RGB}{230,235,252}
\definecolor{myframetitle}{RGB}{90,89,170}
\definecolor{myblocktitle}{RGB}{140,185,249}
\definecolor{mytitle}{RGB}{10,80,26}

\definecolor{darkgreen}{RGB}{27,130,45}
\definecolor{darkblue}{rgb}{0,0,0.3}
\definecolor{darkred}{rgb}{0.7,0,0}

\definecolor{light gray}{RGB}{220,220,220}
\definecolor{dark purple}{RGB}{108,0,217}
\definecolor{pink}{RGB}{190,20,100}
\definecolor{orang}{RGB}{193,63,0}
\definecolor{green}{RGB}{11,98,17}
\definecolor{darkpink}{RGB}{153,0,76}
\definecolor{bluegreen}{RGB}{0,102,102}
\definecolor{greenlagan}{RGB}{0,102,0}
\definecolor{redgreen}{RGB}{102,102,0}
\definecolor{Redgreen}{RGB}{153,76,0}
\definecolor{vividviolet}{rgb}{0.62, 0.0, 1.0}
\definecolor{amaranth}{rgb}{0.9, 0.17, 0.31}
\definecolor{palatinateblue}{rgb}{0.15, 0.23, 0.89}
\definecolor{brightpink}{rgb}{1.0, 0.0, 0.5}
\definecolor{cornflowerblue}{rgb}{0.39, 0.58, 0.93}
\definecolor{deepcarminepink}{rgb}{0.94, 0.19, 0.22}
\definecolor{radicalred}{rgb}{1.0, 0.21, 0.37}

\DeclareFontFamily{OT1}{rsfs}{}

\DeclareFontShape{OT1}{rsfs}{m}{n}{ <-7> rsfs5 <7-10> rsfs7 <10->rsfs10}{} 

\DeclareMathAlphabet{\mycal}{OT1}{rsfs}{m}{n}

\newcommand{\be}{\begin{equation}}
\newcommand{\ee}{\end{equation}}
\newcommand{\bec}{\begin{center}}
\newcommand{\eec}{\end{center}}

\definecolor{olivegreen}{RGB}{60,128,49}

\newcounter{emequation}

\usepackage[most]{tcolorbox}

\tcbset{highlight math style={left=02mm,right=02mm,top=-1mm,bottom=-1mm}} 
\usepackage{empheq}

\begin{document} 

\title{Prism Effect in Quantum Gravity}

\author{Aliasghar Parvizi}
\email{a.parvizi@ipm.ir}
\affiliation{Faculty of Mathematics and Computer Science,\\
Transilvania University, Iuliu Maniu Str. 50, 500091 Brasov, Romania}

\author{Tomasz Pawłowski}
\email{tomasz.pawlowski@uwr.edu.pl}
\affiliation{University of Wrocław, Faculty of Physics and Astronomy, Institute of Theoretical Physics, Maksa Borna 9, PL-50-204 Wrocław, Poland}

\begin{abstract}
Modifications to the dispersion relation of electromagnetic (EM) waves are a central probe in the search for quantum gravitational effects. In this work, we construct a general framework for the interaction between the EM field and a quantum background geometry, employing an extended Born–Oppenheimer approximation. This leads to a quasi-phenomenological model for EM wave propagation in curved spacetime. Unlike previous semi-classical approaches for mode-dependent dispersion relations, our framework naturally reproduces chromatic dispersion effects analogous to those observed in light–matter interactions in nonlinear optics. As a concrete application, we analyze EM wave propagation on a flat quantum Friedmann–Lemaître–Robertson–Walker (FLRW) background, combining analytical techniques with numerical simulations to extract observable signatures of the prism-like behavior induced by quantum light-geometry interactions. Crucially, it remains valid across all energy regimes, enabling access to quantum gravitational corrections beyond the semi-classical limit.
\end{abstract}

\maketitle 

The idea that the quantum nature of gravitational interactions may modify the way matter propagates over a spacetime (as compared to predictions of General Relativity) has brought a considerable attention in the physics community, as it offers a potential for identifying an observable signature of quantum gravity, especially through the modification of the dispersion relation of EM radiation. This line of inquiry can be traced back to Amelino-Camelia et al.~\cite{Amelino-Camelia:1996bln}, who analyzed wave propagation in string theory. Subsequent work explored quantum deformations of the Poincaré group \cite{Magueijo:2001cr,Amelino-Camelia:2000stu}, with phenomenological implications further examined in, e.g. \cite{Amelino-Camelia:1997ieq, Amelino-Camelia:2008aez, Gambini:1998it, Alfaro:2001rb, Magueijo:2002xx}.

With the development of alternative approaches to quantizing gravity, it became natural to pose a similar question within their context. One of such is loop quantum gravity (LQG) \cite{Thiemann:2007pyv, Rovelli:2004tv}, a non-perturbative approach stemming from the strict imposition of the principle of background-independence, consequently employing an analog of a Weyl algebra of variables --holonomies of the gravitational connection and fluxes of the triad--  and a non-standard (known as polymeric) quantum representation. LQG predicts a discrete structure of spacetime at the Planck scale, which manifests itself through discrete spectra of geometric operators (i.e., areas, volumes). 
Loop Quantum Cosmology (LQC) \cite{Ashtekar:2011ni} adapts LQG methods to cosmological settings, focusing on homogeneous and isotropic models of the universe. One of its key features is the resolution to the classical big bang singularity through a bounce mechanism \cite{Ashtekar:2006rx}, which is a consequence of the discrete nature of geometry.
A particularly promising approach to studying the propagation of EM field in quantum spacetime, rooted in the techniques and results of LQC, was first introduced by Ashtekar et al. in \cite{Ashtekar:2009mb} and extended in \cite{Lewandowski:2017cvz, Parvizi:2021ekr}.

In this work, we present a general framework for the interaction of EM field with the quantum geometry. EM fields induce subtle alterations in the quantum state of geometry (the ``atoms'' of spacetime). Treating these fields as perturbations, we employ a perturbative approach and apply time-independent response theory, commonly used in nonlinear optics \cite{Boyd:2008eba}, to analyze their effects on the system's wavefunction and energy. 
These corrections are analogous to the wavelength dependence of linear susceptibilities observed in electromagnetic propagation through solids.
In the context of light-geometry interactions, we draw an analogy to chromatic dispersion in light-matter systems, a ``quantum gravity chromatic aberration''. 

We start with the following action
\begin{equation}
S\ =\  \int d^4x  \left[\frac{\sqrt{-g}}{16\pi G}\mathcal{R} -\frac{1}{2} \sqrt{-g} \rho_T \left( g^{\mu \nu} \partial_\mu T \partial_\nu T + 1 \right)\right] + S_{\rm EM}\ ,
\label{action-tot}
\end{equation}
where, $\mathcal{R}$ denotes the gravitational Lagrangian, $T$ represents the irrotational dust field, $\rho_T$ is a Lagrange multiplier enforcing the condition that the gradient of the dust field is timelike, and $S_{\rm EM}$ corresponds to the action of the EM field. Applying the canonical formalism yields a constrained system in which the Hamiltonian (scalar) constraint takes the form
\begin{equation}
H\ =\ \int d^3x \big[\mathcal{H}_{\rm gr} + \mathcal{H}_{\rm T} +\mathcal{H}_{\rm EM}\big] = 0,
\label{Ham-tot}
\end{equation}
where $\mathcal{H}_{\rm gr}$, $\mathcal{H}_{\rm T}$, and $\mathcal{H}_{\rm EM}$ are the Hamiltonian densities of gravity, dust, and EM field sectors, respectively. In this system, we have $\mathcal{H}_{\rm T} = p_T$, where $p_T$ is the momentum conjugate to the dust field $T$, satisfying $\{T, p_T\} = 1$ \cite{Husain:2011tk, Brown:1994py}. Thus, the system can be deparametrized so that $T$ serves as the evolution parameter. By choosing the gauge $T = t$, time is directly identified with the dust field, and the dynamics is then governed by the (deparametrized) true Hamiltonian $\mathbb{H}$,
\begin{equation}\label{trueH}
    \mathbb{H} \equiv -p_T = H_{gr}(q, \pi) + H_{\rm EM}(Q,P;q),
\end{equation}
with respect to the new time variable $T$. Here, the pair $(q, \pi)$ denotes gravitational canonical gauge-invariant variables, which can be related to the three-metric and its conjugate momenta, and are appropriate for the chosen quantization scheme for gravity. For a more detailed analysis of such systems, see \cite{Giesel:2007wi,Husain:2011tk}. Here, we employ a general framework to extract the phenomenological observables.
The pair $(Q, P)$ represents the EM canonical variables. 
These canonical pairs generate the following Poisson brackets,
\begin{equation}\label{rel:algebra}
    \{q_{ij}, \pi^{kl}\}=\delta_{(i}^k \delta_{j)}^l \qquad \text{and} \qquad \{Q, P\}= 1 .
\end{equation}
The development presented here integrates two key aspects: (i) natural time gauge fixing, and (ii) simplification of the physical Hamiltonian by using irrotational dust field. Together, these elements contribute to a streamlined framework for the quantization of different sectors of the theory.
We assume the existence of a well-defined quantum theory for the system under consideration. More specifically, we postulate that the algebra \eqref{rel:algebra} admits a proper representation on a suitably chosen Hilbert space. Following the Dirac quantization procedure, the total kinematical Hilbert space associated with gravity, matter, and dust system can be constructed as $\mathcal{H}_{\rm kin} = \mathcal{H}_{\rm gr} \otimes \mathcal{H}_{\rm EM} \otimes \mathcal{H}_{\rm T}$. In any canonical framework, spatial coordinate independence is ensured by imposing the spatial diffeomorphism (momentum) constraint, restricting the states to a spatially invariant subspace. Crucially, by utilizing the dust field as a relational physical clock ($T=t$), time-reparametrization ceases to be a gauge symmetry. Consequently, the Hamiltonian constraint is deparametrized into a true Schrödinger-like evolution equation. This allows the spatially invariant subspace to directly serve as the true physical Hilbert space $\mathcal{H}_{\rm phys}$, upon which the physical Hamiltonian $\mathbb{H}$ operates \cite{Husain:2011tk}. Each term in the true physical Hamiltonian \eqref{trueH} can now be promoted to a tensor product of operators of the form $\hat{O} = \hat{O}_{\rm gr} \otimes \hat{O}_{\rm EM}$, where the gravitational part, $\hat{O}_{\rm gr}$, acts solely on the geometric degrees of freedom within the physical Hilbert space, and the matter part, $\hat{O}_{\rm EM}$, acts exclusively on $\mathscr{H}_{\rm EM}$.

\textit{Analysis of the system} ---
To analyze the interaction between gravitational and EM fields and to understand how the backreaction of EM quantum field on the background spacetime can be computed, we will employ an approximation scheme of Born–Oppenheimer type. 
In molecular physics, the Born-Oppenheimer (BO) approximation assumes that the wave functions of atomic nuclei and electrons in a molecule can be treated separately. This is based on the fact that nuclei are significantly heavier than electrons due to their larger relative mass. Consequently, we can solve the eigenvalue problem of the entire system perturbatively and compute the backreaction effects \cite{Kiefer:2004xyv}.
For such an approximation scheme to be viable in the context of quantum gravity, one must assume that the geometrical variables $q$ are quantized as multiplication operators rather than as derivative operators \cite{Giesel:2009at}.
We will operate at the linear order and compute the back-reaction of EM field within the BO approximation for the system \eqref{trueH}. 
To initiate the BO approximation for the gravity plus EM system, we assume a separation of energy scales, treating the EM field as the light degrees of freedom and the geometry states as the heavy one. The total wave function of the light-geometry system is then approximated by the ansatz: $\Psi(q, Q, T) = \psi(q, T) \otimes \phi(Q, T;q)$, where $\psi(q, T)$ and $\phi(Q, T;q)$ are the wave functions of geometry and EM modes, respectively. 
The EM Hamiltonian, $ \hat{H}_{\rm EM}(Q,P;q)$, is intricately entwined with geometry operators.
In general, we plan to perform a spectral decomposition of the Hamiltonian of the system by solving the following eigenvalue problem
\begin{equation}\label{eq:fulleigenf}
    \hat{\mathbb{H}}(\hat{q}, \hat{\pi}, \hat{{Q}}, \hat{{P}}) \, \Psi^\mu(q,{Q}) = E^\mu \, \Psi^\mu(q,{Q}) 
\end{equation}
where $\mu$ is labeling the eigenvalues $ E^\mu$ and the corresponding eigenfunctions $\Psi^\mu$. Solving it exactly is extremely challenging, however, under certain assumptions, it is possible to obtain approximate solutions: Assuming that (i) there is an energy difference and (ii) we have some control over one half of
the problem, which means we can solve or have some knowledge on the eigenvalue problem for the light variables
\begin{equation}\label{eq:eigenfield}
     \hat{H}_{\rm EM}(\hat{Q},\hat{P};q) \phi^n (Q;q) = e^n(\lambda_i; q) \, \phi^n (Q;q)
\end{equation}
$\hat{H}_{\rm EM}$ is an operator acting on the space $\mathscr{H}_{\rm EM}$ that depends on an external parameter $q$, and $\phi^n (Q;q)$ (labeled by $n$) are square integrable in $\mathscr{H}_{\rm EM}$. For the Born–Oppenheimer approximation to work, the dependence of $\phi(Q, T;q)$ on $q$ should be interpreted in a parametric manner and regard \(\phi(Q,T;q)\) as an element of the space \(\mathscr{H}_{\rm EM}\) 
\cite{Giesel:2009at}. As long as the system exhibits a clear separation of energy scales \(\bigl(e^{n}(\lambda_i; q) \ll E^{\mu}\bigr)\), and we work in a representation in which \(q\) acts as a multiplication operator rather than a derivative operator, this approximation remains valid and \(q\) may be replaced by its classical (expectation) value. 
Thus, we can solve \eqref{eq:eigenfield} for each $q$ separately, assuming that the change in $q$ is slow, under the assumption of adiabaticity for heavy degrees of freedom and obtain eigenvalues $e^n(\lambda_i; q)$ that depend on the external parameter $q$ and internal parameter $\lambda_i$. 
\textit{Global Symmetries} define internal parameters $\lambda_i$ that are essential for defining a set of mode functions guided by the symmetries of the spacetime. For example, if there is a spacetime metric $g_{\mu \nu}$ with spacelike killing vectors, the internal parameters $\lambda_i$ will be the wave number $\mathbf{k}$. 
In order to obtain the full wave functions $\Psi^\mu(q,{Q})$, we take the scalar product of Eq.~\eqref{eq:fulleigenf} with respect to $\phi^n(Q;q)$ in the EM Hilbert space $\mathscr{H}_{\rm EM}$, resulting in the following eigenvalue equation for geometry:
\begin{equation}\label{eq:H-corr}
     [\hat{H}_{\rm gr} + e^n(\lambda_i;\hat{q})] \xi^\mu_n (q) = E^\mu \xi^\mu_n(q) \ , 
\end{equation}
where we used BO approximation ansatz  $ \Psi^\mu(q,{Q}) = \sum_n \xi^\mu_n (q) \otimes \phi^n (Q;q)$ (Assuming that $\phi^n(Q;q)$ forms an orthonormal basis of $\mathscr{H}_{\rm EM}$).
Equation \eqref{eq:H-corr} contains a backreaction term  $e^n(\lambda_i;\hat{q})$ that modifies the eigenvalue problem of the geometry. 
To derive the eigenvalue equation \eqref{eq:H-corr}, we disregard all off-diagonal terms in the matrix elements of $\hat{H}_{\rm gr}$ resulting from the action of the geometry momentum operator $\hat{\pi}$ on $\phi^n(Q;q)$. This simplification holds under the Born–Oppenheimer approximation, which assumes a clear separation of characteristic energy scales $| e^n(\lambda_i;q) - e^m(\lambda_i;q) | \gg |E^\mu - E^\nu|$.  Essentially, this means that the impact of light degrees of freedom on the dynamics of heavy sector is effectively captured solely by their energy eigenvalues. 

\textit{Dressed Metric Approach} ---
Although a classical spacetime is absent at the quantum level, the Schrödinger-like equation \eqref{eq:eigenfield} for the EM field can be interpreted as a quantum theory of matter degrees of freedom on a effective background, whose classical limit corresponds to quantum field theory on curved spacetime \cite{Ashtekar:2009mb,Dapor:2012jg}. Equation \eqref{eq:eigenfield} governs the evolution of the quantum state of the mode $\phi^n(Q;q)$, yet the background geometry is neither fully classical nor entirely quantum \cite{Dapor:2013pka}. It means electromagnetic mode acquires information about the geometry only through the expectation values of the geometric operators $\hat{q}$. This naturally raises the question of whether there exists an effective classical spacetime in which the Schrödinger equation \eqref{eq:eigenfield} admits an equivalent description.
To answer this question, we begin by comparing the Hamiltonian on the left-hand side of the evolution equation \eqref{eq:eigenfield} with the Hamiltonian of the EM field on a classical spacetime for modes $\phi^n$. One can assign a dressed spacetime metric $d\tilde{s}^2 = \tilde{g}_{ab}\, dx^a dx^b$ to the evolution of each mode of the EM field on the quantum geometry within the BO approximation. This comparison yields a system of equations that determines the components of $\tilde{g}_{ab}$ in terms of expectation values of various powers of the geometric operator $\hat{q}$. 

To analyze the dispersion relation as an observable, we consider the on-shell condition for a photon with four-momentum \( k_a = (k_0, k_1, k_2, k_3) \), given by $\tilde{g}^{ab}k_a k_b  = 0$. Using the ADM decomposition of the metric, this condition yields
\begin{equation}
\tilde{\omega}(k) = k_0 = \tilde{N}^i k_i + \tilde{N} \sqrt{\tilde{\gamma}^{ij} k_i k_j},
\label{dispersion-classic}
\end{equation}
where \( \tilde{N}^i \), \( \tilde{N} \), and \( \tilde{\gamma}^{ij} \) are the shift vector, lapse function, and spatial metric components of the dressed geometry, respectively.
Similarly, we can define \( \bar{\omega} \), where the metric components are replaced by their expectation values with respect to the unperturbed geometry state, obtained by solving Eq.~\eqref{eq:H-corr} without the backreaction term \( e^n(\lambda_i; q) \). The barred quantities \( \bar{N}^i \), \( \bar{N} \), and \( \bar{\gamma}^{ij} \) thus characterize the unperturbed dressed background, constructed analogously to the back-reacted dressed metric \( \tilde{g}_{ab} \), but excluding backreaction effects.

\textit{Quantum Gravity Toy Model} ---
The framework presented thus far is general and can be applied to any quantization scheme. As an illustrative example, we now consider a flat, isotropic  Friedmann--Lemaître--Robertson--Walker (FLRW) universe, represented by the metric $ds^2 = -N^2(t)\,dt^2 + a^2(t)\,d\boldsymbol{x}^2$ as the background. We employ LQC to quantize the geometric degrees of freedom (see \cite{Ashtekar:2003hd,Ashtekar:2006wn} and the END MATTER for details of the treatment), while the matter sector is quantized according to the Schrödinger picture \cite{ElizagaNavascues:2016vqw, ElizagaNavascues:2020uyf}. As we focus here on the phenomenological implications, a systematic quantization analysis of this toy model is established in \cite{Parvizi:2024vlp}, submitted jointly. 

By fixing the radiation gauge on the flat FLRW background, the electromagnetic field decouples into two physical polarization modes. Mathematically, these modes evolve as independent massless scalar fields, allowing us to decompose the Hamiltonian into a collection of independent harmonic oscillators \cite{Lewandowski:2017cvz} 
\begin{equation}\label{Hamiltonian-SF1}
H_{\rm EM} = \sum_{\mathbf{k}\in{\cal L}} \sum_{r}^2
H_{\mathbf{k}}^{(r)}\ .
\end{equation}
By restricting the background geometry and dust to be strictly homogeneous, and decomposing the electromagnetic field into physical, transverse gauge--invariant Fourier modes, the spatial diffeomorphism constraint is identically satisfied, leaving only the temporal evolution governed by the physical Hamiltonian \eqref{Ham-tot}.
After quantizing the geometry and matter sectors, we find the following Hamiltonian operator for each mode $\mathbf{k}$ and polarization $(r)$ of the EM field: 
\begin{equation}\label{def:emhamiltonian}
    \hat{H}_{\mathbf{k}}^{(r)} = \frac{1}{2}\Big[\langle\hat{V}^{-1}\rangle_\mathbf{k}\, \big(\hat{P}^{(r)}_\mathbf{k}\big)^2 + \ell^{-4} k^2\, \langle\hat{V}^{\frac{1}{3}}\rangle_\mathbf{k}\, \big(\hat{Q}^{(r)}_{\mathbf{k}}\big)^2\Big] \, .
\end{equation}
Here, $Q_{\mathbf{k}}^{(r)}$ and $P_{\mathbf{k}}^{(r)}$ denote conjugate variables associated with the Fourier expansion coefficients of the EM field and its momentum. They satisfy
$\{Q_{\mathbf{k}}^{(r)}, P_{\mathbf{k}^\prime}^{(r^\prime)}\} 
= \delta_{\mathbf{k}\mathbf{k}^\prime}\,\delta_{rr^\prime},$ where the wave vector is restricted to $\mathbf{k} \in \mathcal{L} := \left(\tfrac{2\pi}{\ell}\,\mathbb{Z}\right)^3,$ with $\mathbb{Z}$ denoting the set of integers. In \eqref{def:emhamiltonian} , $\hat{V}$ is the corresponding geometric operator for the physical volume of the universe defined by $V = \ell^3 a^3$, where $\ell$ is the coordinate length of the comoving fiducial cell, and quantized by LQC techniques. If we put $\hat{H}_{\mathbf{k}}^{(r)}$ in  \eqref{eq:eigenfield}, we obtain $ e^n(\mathbf{k};V) = (n+1/2)\hbar \,\mathbf{k} \, \ell^{-2} \langle\hat{V}^{-1/3}\rangle_\mathbf{k}$, where $n$ is the number of particles with momentum $\mathbf{k}$ \cite{Parvizi:2021ekr}. \(\langle \cdot \rangle_{\mathbf{k}}\) denotes the expectation value with respect 
to the geometry state that incorporates the backreaction of mode \(\mathbf{k}\). 
This state is constructed from the eigenfunctions of Eq.~\eqref{eq:H-corr}.

As outlined earlier, by comparing Hamiltonian 
\eqref{def:emhamiltonian} with the Hamiltonian of the EM field 
on a classical FLRW background, one can assign a dressed spacetime $\tilde{g}_{ab}\,dx^a dx^b = -\tilde{N}^2(T)\,dt^2 + \tilde{a}^2(T)\,d\boldsymbol{x}^2$ to each mode $\mathbf{k}$ and polarization $(r)$ of the EM field. This procedure yields the following relations for the metric components:
\begin{equation}\begin{split}\label{dressedBRB}
&\tilde{N}/\tilde{a}^3 = \ell^{-3} \big\langle\hat{V}^{-1}\big\rangle_\mathbf{k}\ , \quad \tilde{N}\tilde{a} = \ell^{-3} \big\langle\hat{V}^{1/3}\big\rangle_\mathbf{k} , \\
& \tilde{N} = \ell^{-3} \left[\big\langle\hat{V}^{-1}\big\rangle_\mathbf{k} ~\big\langle\hat{V}^{1/3}\big\rangle_\mathbf{k}^3 \right]^{\frac{1}{4}},\quad \tilde{a} = \left[\big\langle\hat{V}^{1/3}\big\rangle_\mathbf{k}\, \big\langle\hat{V}^{-1}\big\rangle_\mathbf{k}^{-1}\right]^{\frac{1}{4}}.
\end{split}\end{equation}

\textit{Group Velocity} ---
In general, a low energy cosmological  observer with a normalized 4-velocity $u^a=(1/\bar{N}_T, 0, 0 ,0)$  measures the energy of a  particle with the 4-momentum $k_a=( k_0, k_1, k_2, k_3) $ to be
$E = k_a u^a$. 
The normalization condition for the 4-velocity implies $\bar{g}_{ab}u^au^b=-1$. 
The appropriate rescaled components of the physical momentum $p$ and the energy $E$ in the tetrad frame of the observer (where $\bar{g}_{ab}=\eta_{AB}e_{a}^{A} e^{B}_{b}$, with the internal indices $A, B=0, 1, 2, 3$ and the internal metric $\eta_{AB}$ such that $\eta^{AB} = e^{A a} e^{B}_{a}$) are given by
\begin{equation}
E :=\ k_{\hat{0}}\ = \  \frac{\tilde{\omega}}{\bar{N}_T}\ =\ \mathcal{F}(k, T) \; p  , \quad {\rm and} \quad k_{I}\ =\ \frac{k_{i}}{\bar{a}} ~, 
\end{equation}
where $k_{A}=e_{A}^{a} k_a $ and $p^2= k_{I} k^{I}$  (with $\hat{0}$ denoting the zeroth component of the internal metric and  $I, J=1, 2, 3$ being the three-dimensional internal  indices). 
Then, the 3-velocity $\mathcal{V}^{I}$ of the photon measured in the three-dimensional internal basis  of the cosmological  observer reads
\begin{equation}\label{eq:groupv}
\mathcal{V}^{I}(k, T) = \frac{dE}{dk_{I}}\ ,
\end{equation}
with the squared norm $|\mathcal{V}|^2=\mathcal{V}_I \mathcal{V}^I$, from which we obtain
\begin{equation}\label{eq:velocity-diff}
\mathcal{V}(k, T) = \mathcal{F}(k, T) \left[ 1 - E \frac{\mathrm{d}}{\mathrm{d}E} \log \mathcal{F}(k, T)  \right] \, .
\end{equation}
In our toy model, using the dressed metric components \eqref{dressedBRB}, the dispersion $\tilde{\omega}(k)$ takes the following form
\begin{equation}\begin{split}
\tilde{\omega}(k) &= \dfrac{\tilde{N}}{\tilde{a}} k = k \left( \langle\hat{V^{-1}}\rangle_\mathbf{k} \; \langle\hat{V^{1/3}}\rangle_\mathbf{k} \right)^{1/2}   \\
&= \Big( \frac{\langle\hat{V^{-1}}\rangle_\mathbf{k} \; \langle\hat{V^{1/3}}\rangle_\mathbf{k}}{\langle\hat{V^{-1}}\rangle_o \; \langle\hat{V^{1/3}}\rangle_o} \Big)^{(\frac{1}{2})} \;\bar{\omega} =: \mathcal{F}(k, T)\;\bar{\omega} \, .\quad 
\label{dispersion-BO}
\end{split}\end{equation}
where we defined $\bar{\omega} = (\langle\hat{V}^{-1}\rangle_o \langle\hat{V}^{1/3}\rangle_o)^{1/2} k =: {\bar{N}}/{\bar{a}} \, k$, where $\langle \cdot \rangle_o$ represents the expectation value with respect to the unperturbed geometry state obtained by solving Eq.~\eqref{eq:H-corr} without the backreaction term $e^n(\mathbf{k};V)$. The function $\mathcal{F}(k, T)$ then measures the deviation introduced by backreaction relative to this reference background $\bar{g}_{ab}$.

To analyze the macroscopic implications of this quantum backreaction, we can transition to a semiclassical \textit{effective dynamics} framework. Applying the Ehrenfest theorem to the quantum Hamiltonian \eqref{trueH} for our cosmological toy model, directly relates the expectation values of geometric operators to the classical phase space variables. This yields a modified Friedmann equation for the scale factor $\hat{a}$:
\begin{equation}\label{eq:LQC-Eff-br}
    \begin{split}
\Big( \frac{\overset{\boldsymbol{\cdot}}{\langle \hat{a} \rangle}}{\langle\hat{a}\rangle} \Big)^2
    =\ \frac{8\pi G}{3 }\ \Big( \varrho_{dust} + \frac{\beta \hbar k}{\ell^6 \langle\hat{a}\rangle^{4}}  \Big)
    \times \Big[1-\dfrac{2\alpha_o^2}{3\pi G }\ \Big( \varrho_{dust} + \frac{\beta \hbar k }{\ell^6 \langle\hat{a}\rangle^{4}} \Big) \Big] .
    \end{split}
\end{equation}
The dot indicates differentiation with respect to the internal time $T$, and we used: $\varrho_{dust} := p_T /\ell^3 \langle\hat{a}\rangle^3$ and $\beta:=(1/2+n)$.
For comparison, we also consider the geometrodynamics framework, wherein the Wheeler–DeWitt (WDW) equation and its corresponding phase space variables are employed to derive the evolution equation in the semiclassical (large volume) limit. The analogous effective dynamics for WDW can then be obtained by setting $\alpha_o = 0$ in \eqref{eq:LQC-Eff-br} (where $\alpha_o$ is LQC parameter).

In our evaluations, the dispersion relation coefficients were determined in four regimes:
\begin{enumerate}[(i)] 
  \item \label{it:fullLQC} {\it genuine quantum LQC framework}, where $\mathcal{F}$ has been evaluated via full expression defined in \eqref{dispersion-BO}, with the expectation values determined numerically from a class of semiclassical quantum states, constructed from eigenfunctions of \eqref{eq:H-corr};
  \item \label{it:hybridLQC} {\it hybrid LQC framework}, where the approximation $\langle\hat{V}^n\rangle \approx \langle\hat{V}\rangle^n$ is employed, resulting in the simplified formula $\mathcal{F}(k, T) = \Big( \langle\hat{v}\rangle_o / \langle\hat{v}\rangle_k \Big)^{1/3}$,
  \item \label{it:effLQC} {\it effective LQC framework}, where the trajectories for geometry variables have been determined via Eq.~\eqref{eq:LQC-Eff-br} and applied to evaluate $\mathcal{F}(k, T)$ and $\mathcal{V}(k, T)$,
  \item \label{it:effWDW} {\it effective geometrodynamics framework}, where Eq.~\eqref{eq:LQC-Eff-br} with $\alpha_o = 0$ is employed, which effectively reproduces general relativity (GR) dynamics. 
\end{enumerate}
The results of the genuine quantum and hybrid frameworks were subsequently compared with the semiclassical effective descriptions of both LQC and geometrodynamics, where we assume $ V :=\langle \hat{V} \rangle \approx \ell^3 a^3(T)$ and $\langle \hat{a} \rangle \approx a(T)$. 
Regimes \eqref{it:fullLQC}--\eqref{it:effLQC} allow us to test the accuracy of our simplifications and identify the origins of specific effects, while regime \eqref{it:effWDW} provides a reference point for distinguishing the effects of the loop representation from the standard quantum mechanical treatment employed in the WDW framework. The effective equation \eqref{eq:LQC-Eff-br} can be solved analytically for the WDW regime \ref{it:effWDW}, yielding
\begin{align}
    \mathcal{F}_{\rm WDW}(k, x) &= \sqrt{1 + \epsilon}(1-2 \epsilon) + 2 \epsilon^{3/2} \label{eq:Fkx} \, ,\\
    \mathcal{V}_{\rm WDW}(k, x) &= \mathcal{F}_{\rm WDW}(k, x)\left[1-\frac{\epsilon}{(\sqrt{1+\epsilon} + \sqrt{\epsilon})^2} \right] \, . \label{eq:Vkx}
\end{align}
where $\epsilon = \ell^{-2} \beta\hbar k x/p_T \ll 1$ and $V = x^{-3}$.

In the LQC framework, calculations within the (\ref{it:fullLQC})-(\ref{it:hybridLQC}) regimes were performed using a population of states peaked around $p_T$ ranging from $(500 - 5 \times 10^3) \hbar/\sqrt{G}$, with a relative variance of $\Delta p_T/p_T\in [0.05,0.1]$, a number of EM particles from $1$ to $10$, and a mode index from $25$ to $125$. These expectation values are evaluated using highly peaked semiclassical Gaussian states. The stability analysis, convergence criteria, and operator variance scaling laws for this numerical integration are extensively documented in the companion paper \cite{Parvizi:2024vlp}. Here, we utilize this framework to extract the phenomenological observables. Unlike the WDW trajectory, which reaches a singularity, LQC dynamics reaches the minimum volume,
\begin{equation}\label{eq:Vmin}
     V_{\rm min} \approx \frac{2\alpha_o^2 p_T}{3\pi G} \left( 1 + \frac{\beta \hbar k}{\ell^2 p_T} \, \left( \frac{2\alpha_o^2 p_T}{3\pi G} \right)^{-1/3} \right) \, .
\end{equation}
Notice that the correction term is strictly positive (since $\beta, k, p_T > 0$). This means the electromagnetic backreaction forces the universe to bounce at a slightly LARGER volume than it would if it only contained dust.
\begin{figure*}[tbh!]
    \centering
    \captionsetup[subfigure]{skip=-2pt} 
    
    \begin{subfigure}[b]{0.3\textwidth}
        \centering
        \includegraphics[width=\textwidth]{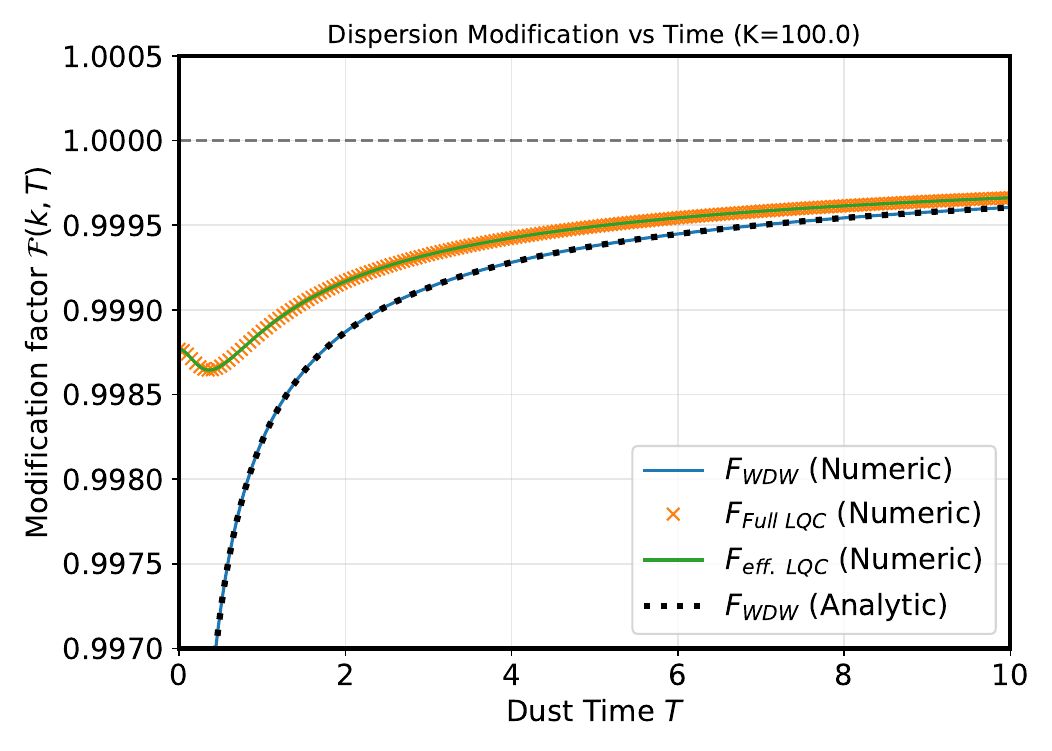}
        \caption{}
        \label{fig:Ft}
    \end{subfigure}
    \begin{subfigure}[b]{0.3\textwidth}
        \centering
        \includegraphics[width=\textwidth]{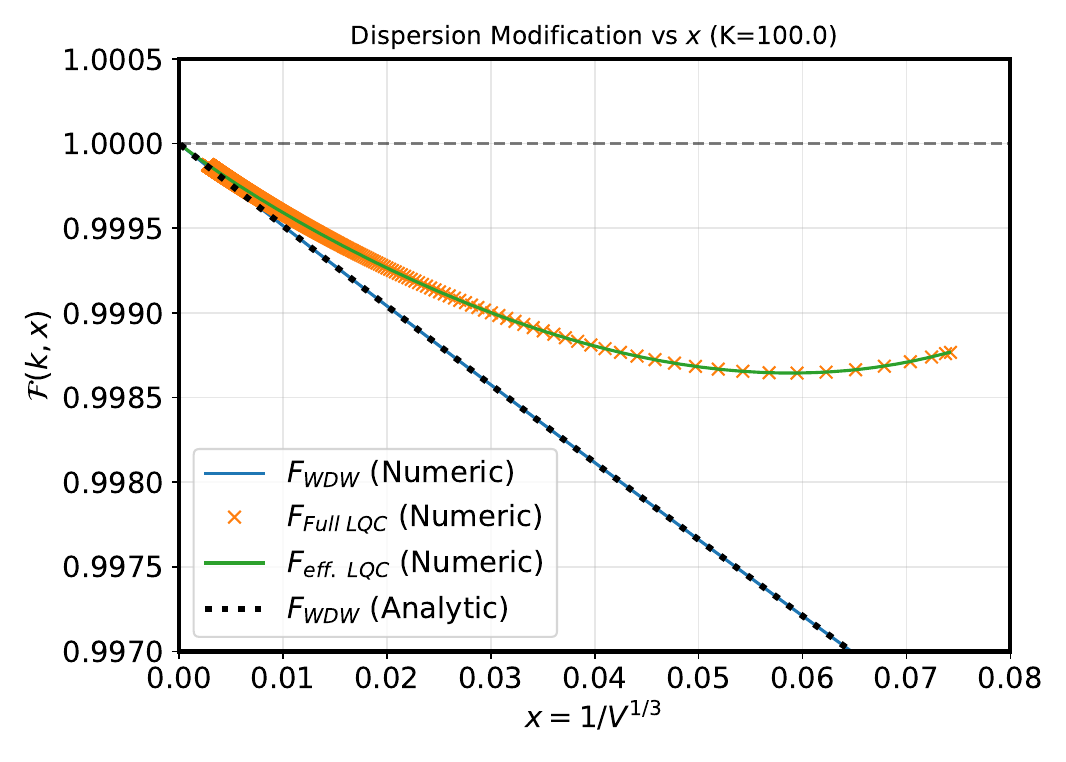}
        \caption{}
        \label{fig:Fx}
    \end{subfigure}
    \begin{subfigure}[b]{0.3\textwidth}
        \centering
        \includegraphics[width=\textwidth]{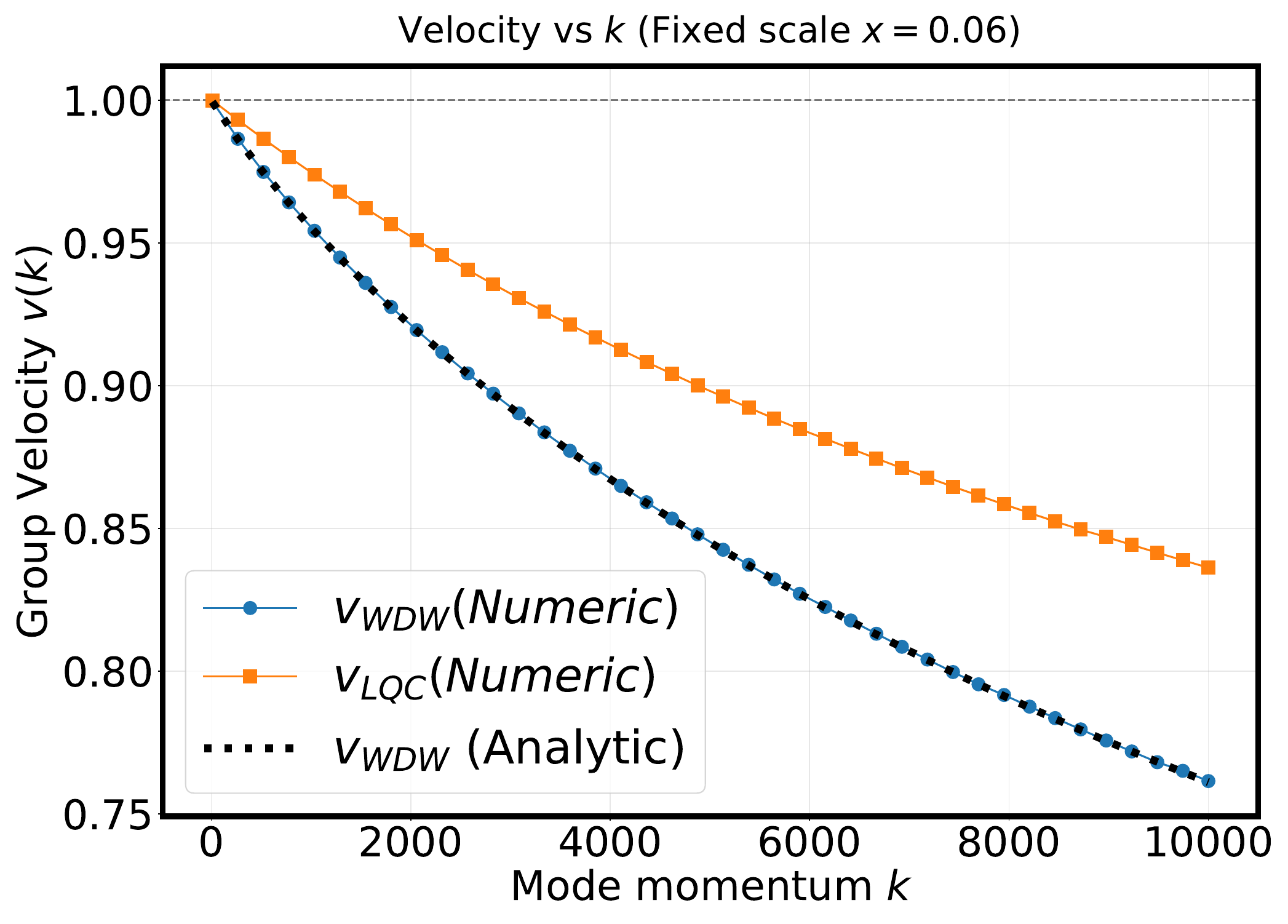}
        \caption{}
        \label{fig:v(k)}
    \end{subfigure}
    
    \begin{subfigure}[b]{0.3\textwidth}
        \centering
        \includegraphics[width=\textwidth]{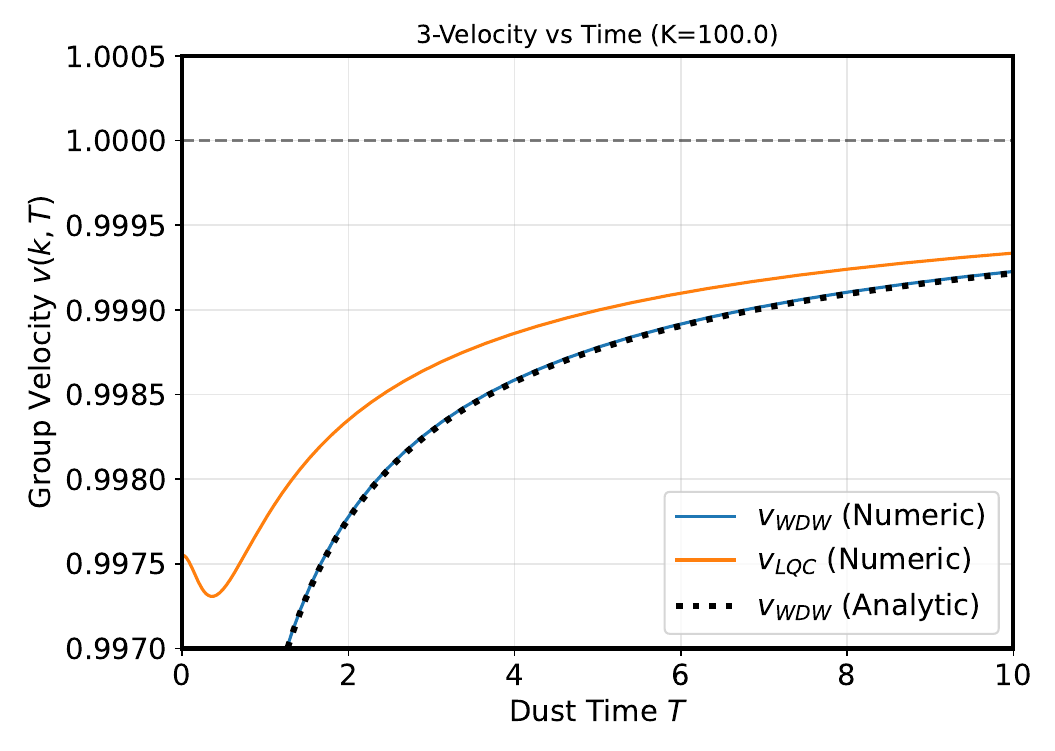}
        \caption{}
        \label{fig:vt}
    \end{subfigure}
    \begin{subfigure}[b]{0.3\textwidth}
        \centering
        \includegraphics[width=\textwidth]{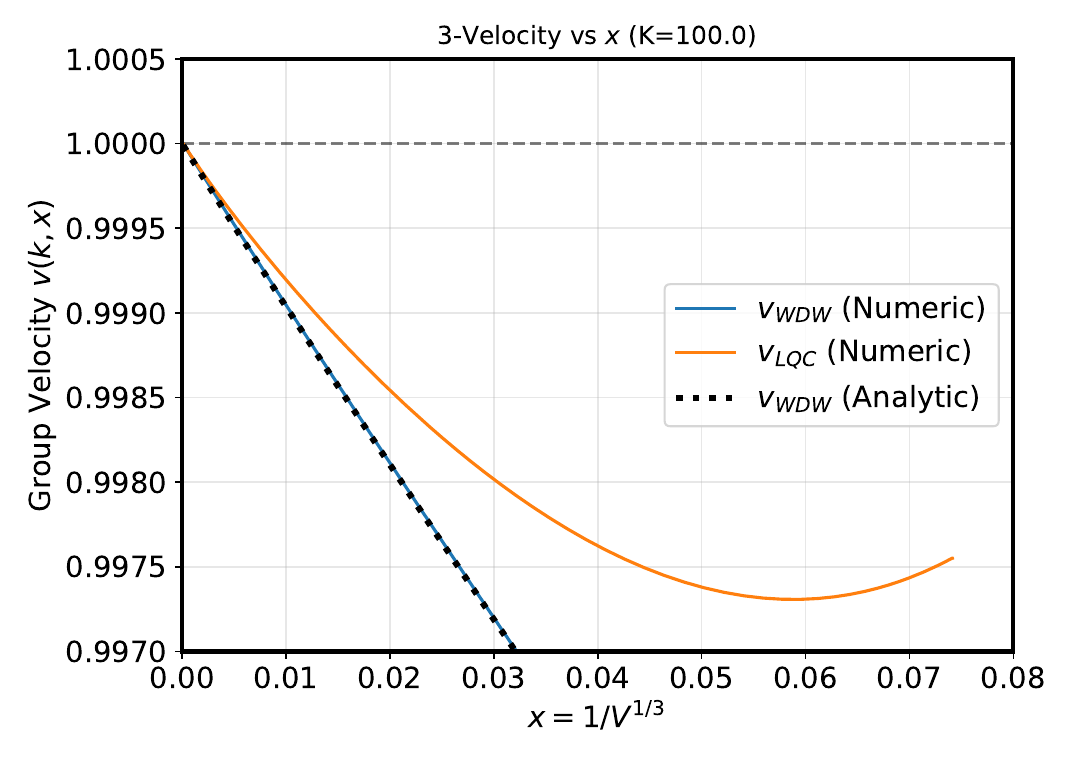}
        \caption{}
        \label{fig:vx}
    \end{subfigure}

    \caption{Dispersion relation coefficient $\mathcal{F}$ and group velocity $\mathcal{V}$ evaluated via methods \eqref{it:fullLQC}-\eqref{it:effWDW} for a Gaussian state peaked about $p_T=10^3\hbar/\sqrt{G}$ with relative variation $\Delta p_T/p_T = 0.05$ with a single particle mode of $k=100$ plotted as function of dust time $T$ and $x=V^{-1/3}$. Bounce point (minimum volume \eqref{eq:Vmin}) happens at $T = 0$ and $x \approx 0.074$ for (a)-(b) and (d)-(e). (c) evaluated at $x = 1/V^{1/3} = 0.06$ for $k \in [10, 10^4 ] \Rightarrow V_{\rm min} \in [0.066,0.074]$.}
    \label{fig:FV-main}
\end{figure*}
An example of the results is presented in Fig.~\ref{fig:FV-main}. The following properties, which are common for the entire population of cases, were observed:
\begin{enumerate}
  \item \label{it:res-traj} The results from the genuine quantum regime~(\ref{it:fullLQC}), \(\mathcal{F}_{\rm Full~LQC}\), and the effective regime~(\ref{it:effLQC}), \(\mathcal{F}_{\rm eff.\,LQC}\), exhibit close agreement in Figs.~\ref{fig:Ft} and \ref{fig:Fx} (see also END MATTER for the relative difference between the genuine quantum (\ref{it:fullLQC}) and hybrid approach (\ref{it:hybridLQC})).
  %
  \item \label{it:res-bound} Fig.~\ref{fig:FV-main} demonstrates that the coefficient $\mathcal{F}$ and velocity $\mathcal{V}$ satisfy the bounds $\mathcal{F}_{\rm WDW} < \mathcal{F}_{\rm LQC} < 1$ and $\mathcal{V}_{\rm WDW} < \mathcal{V}_{\rm LQC} < 1$ for all methods \eqref{it:fullLQC}--\eqref{it:effWDW} within their respective domains of applicability. 
  \item \label{it:res-turn} As demonstrated in Figure~\ref{fig:vt} and \ref{fig:vx}, contrary to the behavior of $\mathcal{V}_{\rm WDW}$, the deviations of $\mathcal{V}_{\rm LQC}$ from unity do not attain a maximum at the bounce point. 
\end{enumerate}
Although the effective dynamics are exact, the full quantum results were obtained numerically only within a restricted domain, primarily for analyzing $\mathcal{F}$, due to computational cost.

To explicitly verify causality, we must analyze the group velocity $\mathcal{V} = |\mathcal{V}^I|$ in the frame of a cosmological observer \eqref{eq:groupv} in low and high energy regimes.
Equation \eqref{eq:Vkx} together with Fig.~\ref{fig:FV-main} highlight two crucial phenomenological features: 
i) In the low-energy limit ($k \to 0$), both the WDW and LQC trajectories converge smoothly and exactly to $\mathcal{V} \approx 1 - 2 \, \epsilon$. This demonstrates that the standard, Lorentz-invariant dispersion relation of General Relativity is restored at low energies;
ii) As the mode momentum $k$ increases, the group velocity acquires a strictly negative correction ($\partial \mathcal{V}/\partial k < 0$, see Fig.~\ref{fig:v(k)}). This guaranties that the chromatic dispersion (the \textit{prism} effect) induced by the quantum geometry is entirely subluminal across all energy scales.

We conclude by placing the presented results in a broader context. We aimed to establish a more robust theoretical and numerical framework to investigate the prism effect in a quantum gravity system as a possible effect of the quantum nature of geometry. 
Our study reveals the following qualitative behaviors for the propagation of the electromagnetic field: (a) the group velocity is mode-dependent, meaning each mode propagates at a different speed; (b) the modification in $\mathcal{V}$ depends on the number of particles $\beta$ in that mode, with this parameter serving as amplification factor for subtle quantum effects ( a quantum gravitational analog to the \textbf{Optical Kerr} effect); (c) the standard dispersion relation is recovered in the low-energy limit; (d) the modification leads to subluminal propagation; and (e) for the model derived from LQC, the modifications to the EM field’s propagation velocity remain bounded throughout the entire cosmic evolution and are smaller than those predicted by the corresponding geometrodynamics model.

We chose the loop quantum gravity framework for our analysis due to its suitability for theoretical and numerical calculations. The underlying ingredients of the construction, such as gauge fixing, using the dust field as an internal time parameter, applying the Born–Oppenheimer approximation, and assigning a dressed background to the quantum evolution, are not tied to any specific quantization scheme of gravity. Consequently, the same procedure for obtaining \(\mathcal{F}\) and \(\mathcal{V}\) can be carried over to other quantum gravity frameworks (including, for example, the geometrodynamics model used here to isolate effects characteristic of loop quantization). However, the qualitative behavior of $\mathcal{F}$ and $\mathcal{V}$ in the high-energy limit, strongly depends on the quantization scheme and particular form of $\hat{H}_{\rm gr}$ (Fig.~\ref{fig:FV-main}). 

The standard phenomenological models of Lorentz Invariance Violation (LIV) have the form: $E^2  = p^2 + \mathcal{A}_\alpha \, p^\alpha $, where $\mathcal{A}$ and $\alpha$ are two parameters that characterize the GR deviation \cite{Mirshekari:2011yq}. In our setup, the factor \( \epsilon \propto \beta\,\hbar k\, x \) exhibits a distinct structure, reflecting both curvature dependence (through the inverse scale factor \( x \)) and intensity dependence (through the occupation number \( \beta \)).
According to our framework, in the realm of multi-messenger astrophysics \cite{AmelinoCamelia:1997gz, Bernardini:2014oya, Yi:2005ht, Piron:2015wtz}, the propagation speed of EM signal acquires an energy-dependent delay $\Delta t \propto \int (1 - \mathcal{V}(E,T)) \;dT$. While strict bounds on LIV, such as those derived from Fermi-LAT's observation of GRB 090510 \cite{FermiGBMLAT:2009nfe}, restrict standard linear deformations to scales beyond the Planck energy ($E_{QG} > M_{Pl}$), the quantum gravity \textit{chromatic aberration} derived here presents a structurally different profile (see Fig.~\ref{fig:v(k)}). 


\begin{acknowledgments}
    This work was supported in part by the Polish National Center for Science (Narodowe Centrum Nauki -- NCN) grant OPUS 2020/37/B/ST2/03604. The work of AP was supported by a grant of the Transilvania University of Brașov (UNITBV), offered through the Transilvania Fellowship for Postdoctoral Research/Young Researchers program.
\end{acknowledgments}

\bibliographystyle{apsrev4-1}
\bibliography{References}

\clearpage 

\section{ End Matter} 

\setcounter{equation}{0} 
\renewcommand{\theequation}{EM.\arabic{equation}} 

\textit{Gravitational Hamiltonian} ---\label{app:gr-ham}
By imposing the canonical time gauge fixing condition, i.e., $N(t)=1$, the gravitational Hamiltonian for homogeneous and isotropic spacetime $ds^2 = - N^2(t) dt^2 + a^2(t) d\mathbf{x}^2$, reads \cite{Parvizi:2021ekr}
\begin{equation}\label{Hgr}
H_{\rm gr} = \int d^3x \mathcal{H}_{\rm gr} =\  \frac{3\pi G}{2\alpha_o}b^2|v| \ , 
\end{equation}
where the phase space is conveniently coordinatized by a canonically conjugate pair $\{b, v\}=2$, where $|v|=a^3/\alpha_o$  and $b=\gamma(\dot{a}/a)$ are the rescaled volume and Hubble parameter, respectively (where ``dot'' refers here to the dust proper time derivative). The constant $\gamma$ is the so-called Barbero-Immirzi parameter of LQG, while $\alpha_o=2\pi \gamma \sqrt{\Delta}  \;\ell_{\rm Pl}^2$, in which $\Delta \equiv 4 \sqrt{3}\pi \gamma \ell_{\rm Pl}^2$ is the LQC {\em area gap} and $\ell_{\rm Pl}$ is the Planck length \cite{Ashtekar:2006rx} and the sign of $v$ encodes the orientation of the triad. Upon promoting the relevant fundamental components to operators, the oriented volume $\hat{v}$ is related to the physical volume of the universe $V$ by $V = \ell^3 a^3 = \alpha_0 |v|$, and the shift operator $\hat{\mathcal{N}} = \widehat{\exp(ib/2)}$, the gravitational Hamiltonian takes the form \cite{Husain:2011tm}
\begin{equation} \label{Ham_grav}
    \hat{H}_{\rm gr} = \frac{3\pi G}{8\alpha_0} \sqrt{|\hat{v}|} \left(\hat{\mathcal{N}}^2 - \hat{\mathcal{N}}^{-2}\right)^2 \sqrt{|\hat{v}|},
\end{equation}
which represents the symmetrized, UV-complete quantum version of the classical Hamiltonian \eqref{Hgr}.

\begin{figure}[tbh!]
    \psfrag{dF}{$\delta\mathcal{F}(x)$}
    \begin{center}
    \includegraphics[width=0.35\textwidth]{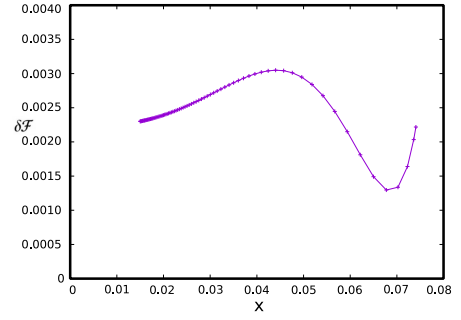}
    \end{center}
    \caption{Relative difference for coefficient $\mathcal{F}$ determined via genuine (\ref{it:fullLQC}) and hybrid method (\ref{it:hybridLQC}) within LQC framework evaluated for the example given in Fig.~\ref{fig:FV-main}, showing a regular stabilization as $x\to 0$.}
    \label{fig:dFx-LQC}
\end{figure}

\textit{Genuine quantum versus hybrid modification for LQC} --- Since in the genuine quantum approach the numerical evaluations are preformed in the volume representation, a direct comparison of the results with the ones in effective (or hybrid) approach is not possible in the low energy limit, as the universe expands to large volumes here, thus imposing unrealistic requirement on the size of the volume domain. Thus, to establish the convergence of the dispersion relation coefficients, one has to resort to an extrapolation. Here, we do so by looking at the relative difference in the modification to the dispersion relation $1-\mathcal{F}$ between the genuine quantum and hybrid approach,
\begin{equation}
    \delta\mathcal{F}(T) = \frac{|\mathcal{F}_q(T) - \mathcal{F}_h(T)|}{1-\mathcal{F}_h(T)} , 
\end{equation}
where $\mathcal{F}_q$ and $\mathcal{F}_h$ are functions $\mathcal{F}$ resulting from the genuine quantum and hybrid approach respectively. In fact, direct inspection of the numerical results shows that $\delta\mathcal{F}$
is bounded within the domain in which it could have been probed (see Fig.~\ref{fig:dFx-LQC}). Furthermore, the shape of $\delta\mathcal{F}$ in the probed region strongly indicates that it will remain bounded also in the low energy limit ($x\to 0$). This includes the regime \(x \in [0,0.01]\) shown in Fig.~\ref{fig:Fx}, which supports the reliability of the results for $\mathcal{V}$ in the effective dynamics.
This means 



\end{document}